\documentstyle[aps,prl,epsf,multicol]{revtex}
\def\myfigure#1#2{{\leftskip=0.000753\textwidth \rightskip\leftskip\small
\begin{figure}\baselineskip=14pt plus 2pt minus 1pt
\centerline{#1}\nobreak\smallskip\nobreak #2\end{figure}}}
\global\firstfigfalse

\begin{document}

\draft

\title{Dynamic scaling in the spatial distribution of persistent sites}

\author{ G.\ Manoj\cite{MAN} and P.\ Ray\cite{RAY}}

\address{ The Institute of Mathematical Sciences, C. I. T. Campus,
Taramani, Madras 600 113, India}

\date{\today}

\maketitle
\begin{abstract}
The spatial distribution of persistent (unvisited)
sites in one dimensional $A+A\to\emptyset$ model is studied. The
`empty interval distribution' $n(k,t)$, which is the probability
that two consecutive persistent sites are separated by distance $k$ 
at time $t$, is investigated in detail. 
It is found that at late times this distribution has the
dynamical scaling form
$n(k,t)\sim t^{-\theta}k^{-\tau}f(k/t^{z})$. The new exponents
${\tau}$ and $z$ change with the
initial particle density $n_{0}$, and are related to the
persistence exponent ${\theta}$ through the scaling relation
$z(2-\tau)=\theta$. We show by rigorous analytic arguments that
for all $n_{0}$, $1< \tau< 2$, which is confirmed by 
numerical results.

\end{abstract}

\begin{multicols}{2}
First passage problems in non-equilibrium systems undergoing
time evolution has become an important field of
research lately with the discovery of persistence. Persistence
probability  in general is defined as follows:
Given a stochastic variable $\phi(t)$ which fluctuates about a mean
value, say zero, 
what is the probability $P(t_{1},t_{2})$ that $\phi(t)$ 
does not change sign throughout the
time interval $[t_{1},t_{2}]$. 
For a large class of physical systems, persistence
shows a power-law decay
$P(t_{1},t_{2})\sim (t_{2}/t_{1})^{-\theta}$ for $t_{2}\gg t_{1}$,
with a non-trivial persistence exponent ${\theta}$
\cite{BRAY,DERRIDA,PHORD,DIFFUSION,GLOBAL,VOTER,HARI,CARDY,BENN,BPLEE,FINT,MOBIL}  
which is, in general, unrelated to other known static and
dynamic exponents.

Let us consider spatially extended systems with a 
stochastic field $\phi({\bf x},t)$ at each lattice site ${\bf x}$,
the time evolution of which is
coupled to that of its neighbouring sites.
$\phi({\bf x},t)$ could be, for instance, an Ising
spin\cite{BRAY,DERRIDA}, a phase ordering
field\cite{PHORD}, a diffusing field\cite{DIFFUSION} 
or the height of a fluctuating interface\cite{FINT}.
In such cases, the system gets broken up 
into domains of persistent and non-persistent sites
in course of time. 
In $d=1$, this reduces to a set of {\it disjoint} persistent and non-persistent
{\it clusters} appearing alternately. 
As persistence decays with time, the persistent clusters
shrink in size and hence, their separation grows.
The following questions arise naturally in this context,
which we address here: (i) How are the persistent 
clusters distributed in space at a given time? (ii) How does
their average separation grow with time? 

In one dimension, the zeroes of the stochastic field can 
be viewed as a set of particles, moving about in the lattice, annihilating each
other when two of them meet. When a particle moves across a lattice
site for the first time, the field there flips sign, and the
site becomes non-persistent. If each particle is assumed to perform
purely diffusive motion, this reduces to the well-known
reaction-diffusion model $A+A\to\emptyset$, with appropriate initial
conditions. The simplest case is random initial distribution of particles,
with average density $n_{0}$, for which $P(t)\sim t^{-\theta}$ 
with ${\theta}=3/8$\cite{DERRIDA}, independent of $n_{0}$
\cite{CARDY}. We investigate		
spatial ordering of persistent sites in this simple model. 

Our study is centered around the Empty Interval Distribution $n(k,t)$ 
--- the probability that two randomly chosen consecutive persistent 
sites are separated
by distance $k$ at time $t$. 
This distribution is analogous to 
the well-studied Inter-Particle Distribution Function (IPDF)
in diffusion-reaction systems\cite{REVIEW}.
Our numerical results show that   
$n(k,t)$ has a non-trivial dynamic scaling form with
{\it power-law decay}
in $k$ and $t$, characterized by exponents ${\tau}$
and ${\theta}$ respectively.
The power law decay is valid    
for $k\ll L(t)$, where $L(t)\sim t^{z}$
is a new dynamic length scale, which may be interpreted as
the average separation between persistent regions.  
The three exponents are connected by the scaling relation
$z(2-\tau)=\theta$. Although the persistence exponent
$\theta$ is universal
for this model, we find that $\tau$ and $z$ do change with the
initial particle density. We give rigorous analytical
arguments on the bounds of ${\tau}$, showing that $1< \tau < 2$
for all values of $n_{0}$. Power-law decay of $n(k,t)$ in
$k$ is a consequence of spatial correlations---
a random distribution of sites would correspond to exponential decay.

Our numerical simulation is done
on a 1-d lattice of size $N= 10^{4}$ with periodic boundary conditions.
Particles are initially distributed at random on the lattice, and
their positions are sequentially updated--- each particle
was made to move one step in either direction with probability 
1/2. When two particles came on top of each other, both vanished
instantaneously. The time evolution is
done up to 12000 Monte-Carlo steps (1 MC step is counted after all the
particles in the lattice were touched once). All simulations are repeated
for three different values of initial density, $n_{0}=0.2$, 0.5 and 0.8. 
The results are averaged over 500 different initial configurations. 

We observe that at large times $t$ and $k\gg 1$,
$n(k,t)\sim k^{-\tau}$ for $k\ll L(t)$. 
Here $L(t)$ is a cut-off length scale that grows with time. 
In Fig. 1, we present the data
for $n_{0}=0.5$ and for three values of time. 
The same data as presented in Fig. 2 shows that for each $k$,
$n_{k}(t)\sim t^{-\omega}$ for {\it all} $k\ll L(t)$.
(It will be shown later that $\omega=\theta$).
Similar power-law decay in $k$ and $t$ has been observed for other
values of $n_{0}$ also.
These observations are fairly well represented by
the following dynamic scaling 
form for $n(k,t)$, for late times and large enough $k$.  

\begin{equation}
n(k,t)\sim t^{-\omega}k^{-\tau}f(k/L(t))
\label{eq:SCALFORM}
\end{equation}

where 
the scaling function $f(x)\simeq 1$ for $x\ll 1$ 
and decreases faster than any power
of $x$ for $x\gg 1$.

\myfigure{\epsfysize3.5in\epsfbox{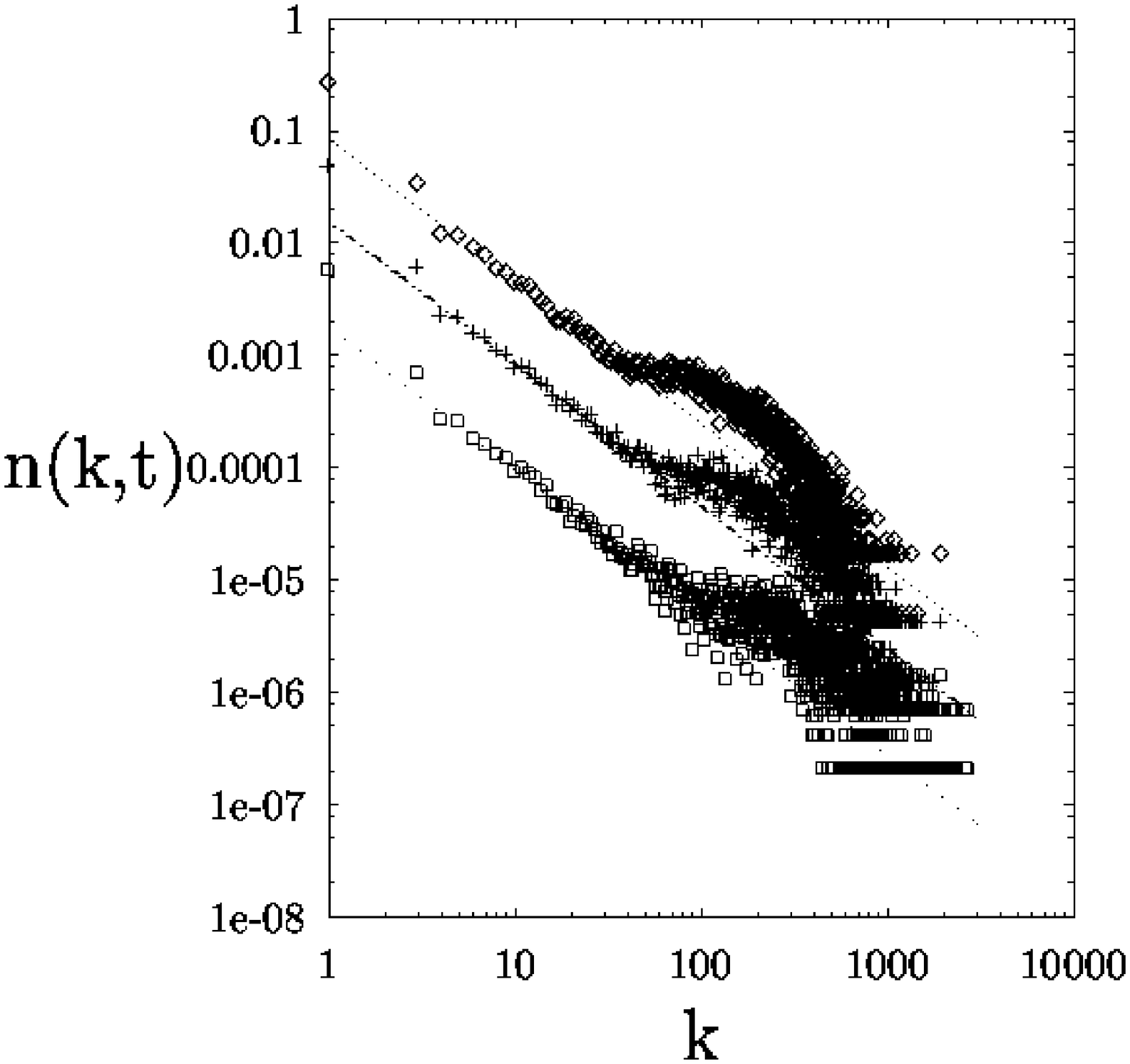}}
{\vskip-1.2in Fig.\ 1~~ 
Log-log plot of $n(k,t)$ with $k$ illustrating the power law decay
for small $k$, which crosses over to faster decay at large $k$.
The data is presented for $t=2000, 4000$ and 10000 (top to bottom). 
The vertical separation between the curves has been enhanced for clarity.
The initial density $n_{0}$ is 0.5.
All the three straight lines have slope ${\tau}\simeq 1.27$.}

\myfigure{\epsfysize2.8in\epsfbox{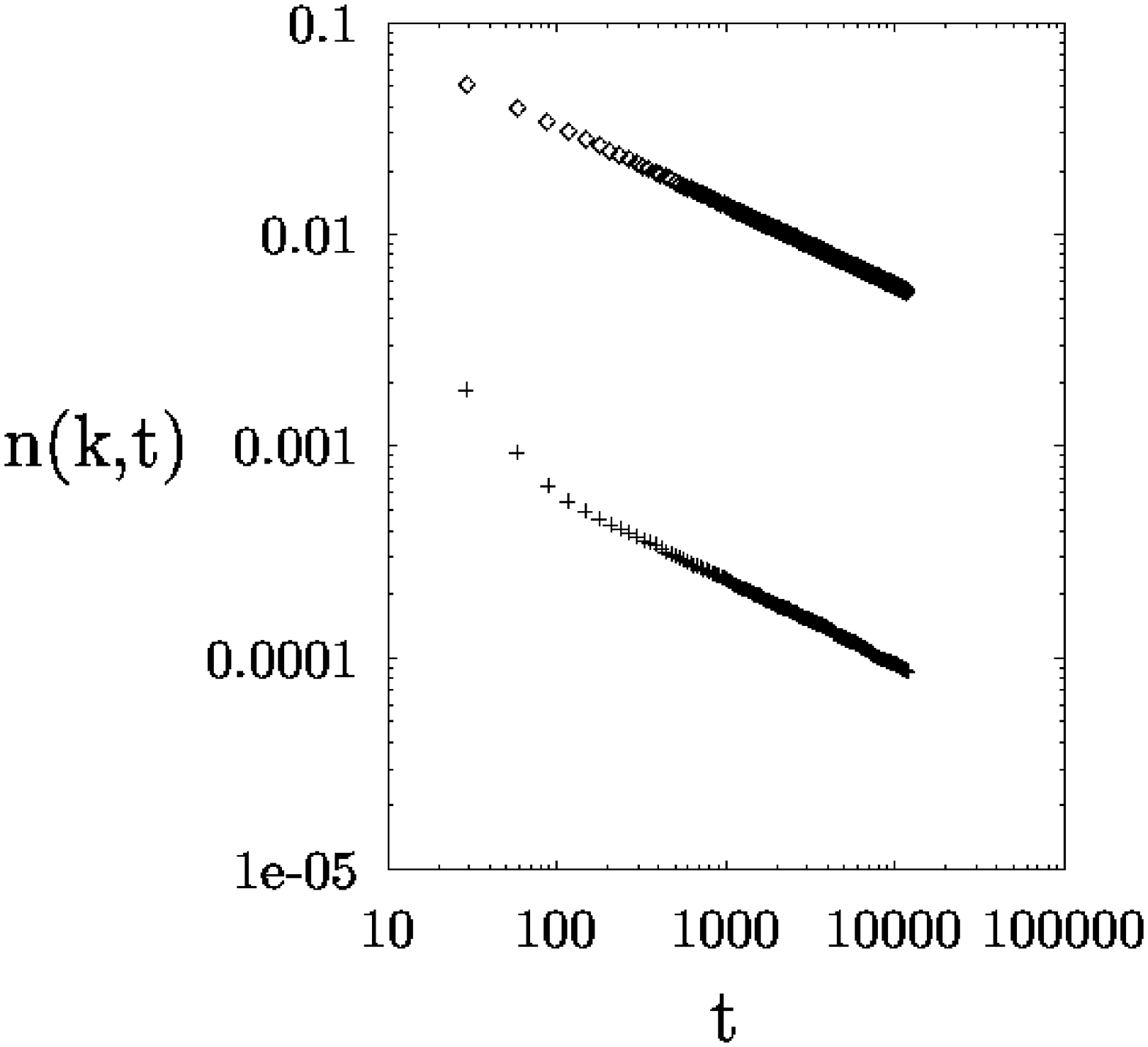}}
{\vskip-1.0in Fig.\ 2~~ 
Log-log plot of $n(k,t)$ with $t$ for 
$k=1$ (top) and $k=10$ (bottom). Initial density
$n_{0}$ is 0.5. The asymptotic slope
of the lines gives ${\omega}\simeq 0.37$.}

The exponents appearing in Eq.\ \ref{eq:SCALFORM} are not all independent.
The moments of the distribution are useful in
deriving the scaling relations between them. 
The $m$-th moment is $I_{m}(t)=
\sum_{k}k^{m}n(k,t)\approx \int_{1}^{\infty}n(s,t)s^{m}ds$.
From the definition of $n(k,t)$, one can easily see that

\begin{equation}
I_{0}(t)=P(t)\sim t^{-\theta}\hspace{0.3cm}; \hspace{0.5cm} I_{1}(t)=N 
\label{eq:NORMAL}
\end{equation}

The average separation between persistent sites is given by 
$I_{2}(t)/I_{1}(t)\sim L(t)$. In Fig. (3) we have $L(t)$
plotted against $t$ on a logarithmic scale for three values
of $n_{0}$. We find that $L(t)$ diverges with time as 
\begin{equation}
L(t) \sim t^z
\label{eq:LTZ}
\end{equation} 
where $z$ is a new dynamic exponent. 

\myfigure{\epsfysize3.5in\epsfbox{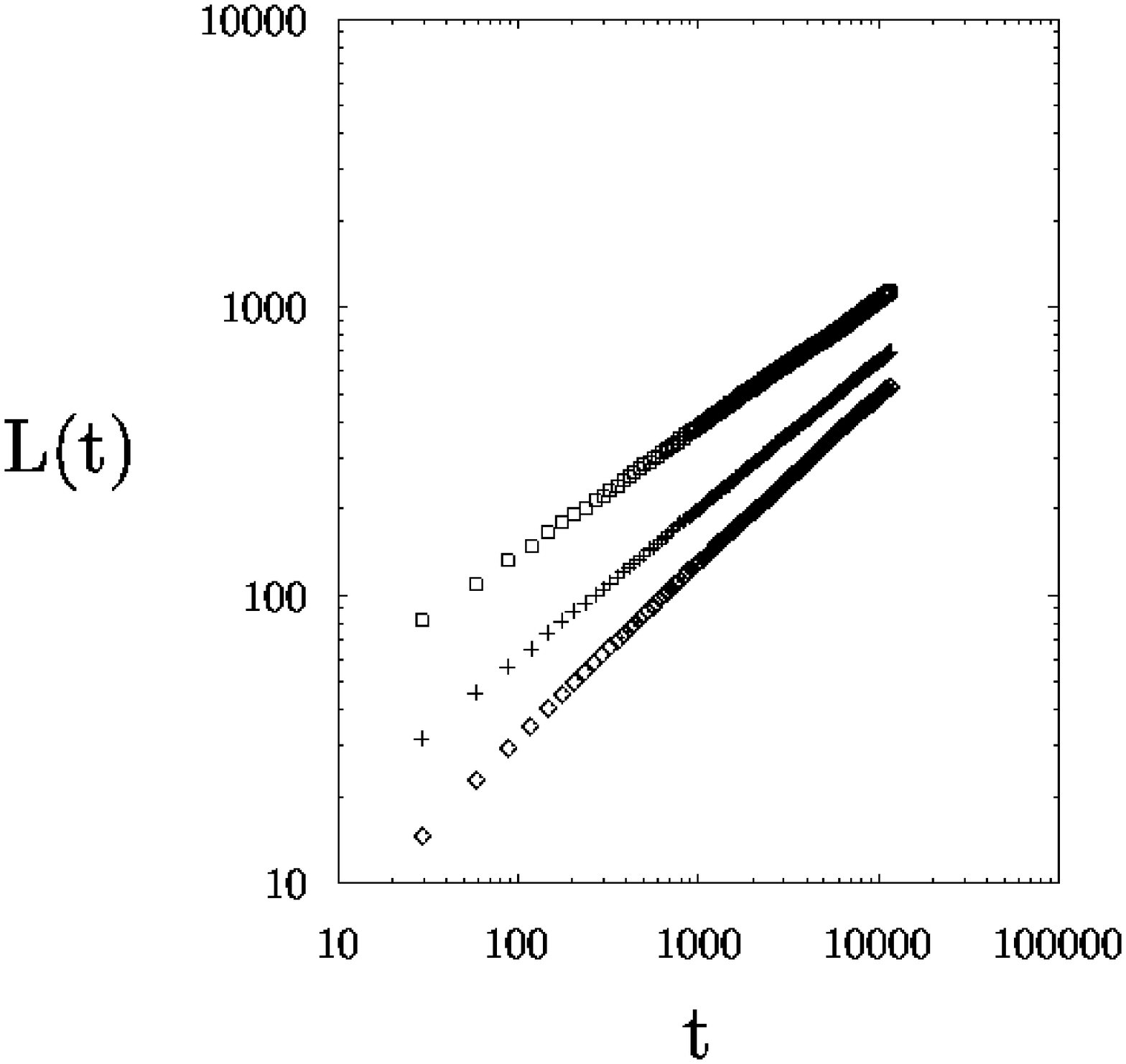}}
{\vskip-1.2in Fig.\ 3~~ 
The average separation between persistent
sites $L(t)$ grows as a power of time $t$. The three logarithmic plots
correspond to  
$n_{0}=0.8$, 0.5 and 0.2 (top to bottom). The lines are visibly
getting flatter with increasing $n_{0}$, indicating the decrease in
the exponent $z$.}

The scaling relations between the 
exponents are obtained by making use of the conditions in Eq.\ \ref{eq:NORMAL}.
First of all, we show that only ${\tau}<2$ is 
physically reasonable. For, if 
${\tau}\geq 2$, $I_{1}(t)\sim t^{-\omega}$,
and from the second part of Eq.\ \ref{eq:NORMAL} it follows that ${\omega}=0$.
But since $\omega\geq\theta$ 
for reasons of convergence, we get ${\theta}\leq 0$
which is absurd. So we conclude that ${\tau}< 2$.
In this case, $I_{1}(t) \sim t^{-\omega +z(2-\tau)}$, 
which according to Eq.\ \ref{eq:NORMAL} imply that

\begin{equation}
z(2-\tau)=\omega
\label{eq:SCAL1}
\end{equation}

Another set of scaling relations can be derived using
the condition on $I_{0}(t)$ in Eq.\ \ref{eq:NORMAL}.  
Combined with Eq.\ \ref{eq:SCAL1}, this gives 

\begin{equation}
z=\theta \hspace{0.3cm}; \hspace{0.5cm} \omega=\theta(2-\tau)
\hspace{0.5cm}$if$\hspace{0.4cm} \tau< 1
\label{eq:SCAL2}
\end{equation}

\begin{equation}
\omega=\theta \hspace{0.3cm}; \hspace{0.5cm} z(2-\tau)=\theta
\hspace{0.5cm}$if$\hspace{0.4cm} \tau> 1
\label{eq:SCAL3}
\end{equation}

We present a summary of our numerical results in Table I. 
It is easily seen that for all values of $n_{0}$,
$z > \theta$, $\tau > 1$ and the scaling relations in
Eq.\ \ref{eq:SCAL3} are satisfied within numerical errors. 
Moreover, the exponents
$z$ and $\tau$ show a consistent decrease
with increasing $n_{0}$--- they are non-universal, unlike $\theta$.
We now present an intuitive argument which accounts for
these observations fairly rigorously.

As persistence decays,
the non-persistent regions grow in time (the length scale of which is
set by $L(t)$ in Eq.\ \ref{eq:SCALFORM}) while the clusters of
persistent sites 
shrink in size and eventually disappear. 
Let $p(l,t)$ be the
number of persistent clusters of           
size $l$, at time $t$. The total number of persistent
sites at time $t$ is $\sum_{l}lp(l,t)=P(t)$, and the
total number of such clusters is $N_{c}(t)=
\sum_{l}p(l,t)$. The latter is related to $n(k,t)$
through the exact relation
$N_{c}(t)=\sum_{k=2}^{\infty}n(k,t)= P(t)-n(1,t)$.
The average size of a cluster at time $t$ is 

\begin{equation}
\overline l(t)
=\frac{P(t)}{N_{c}(t)}=\left(1-\frac{n(1,t)}{P(t)}\right)^{-1}
\label{eq:AV}
\end{equation}

From Eq.\ \ref{eq:SCALFORM}, $n(1,t)\sim t^{-\omega}$ and since 
$P(t)\sim t^{-\theta}$ we have 
$\overline l(t)=\left[1-\gamma t^{-(\omega-\theta)}\right]^{-1}$
where ${\gamma}$ is a numerical constant. Since $\omega \geq\theta$,
$\overline l(t)$ is a constant for late times.
Now, if $\omega> \theta$, $\overline l(t)=1$ strictly;
only if $\omega=\theta$ any other value is possible.
We argue for the latter case as follows.
When clusters of persistent sites shrink in size, the depletion happens 
at the two ends of the cluster, independent of its size. Let the average
decrease in the size of a cluster over time $t$ be $\xi(t)$. Clusters
of initial size $l> \xi(t)$ shrink to size $l-\xi(t)$ after
time $t$, while those with length $l\leq \xi(t)$ disappear. 
It follows that 

\begin{equation}
p(l,t)=p(l+\xi(t),0) 
\label{eq:LPST}
\end{equation} 

Here, $p(l,0)=n_{0}^{2}(1-n_{0})^{l}$ since the initial distribution of
particles is done at random with probability $n_{0}$. After
substitution in Eq.\ \ref{eq:LPST}, we find that the
time evolution of the cluster size distribution 
has the extremely simple form $p(l,t)= e^{-\lambda\xi(t)}p(l,0)$
where $\lambda=-$ln$(1-n_{0})$. 
This result is also supported by simulations (Fig. 4).
Consequently, the average cluster size
$\overline l(t)=\overline l(0)=1/n_{0}$. This implies
$\omega=\theta$ from our arguments following Eq.\ \ref{eq:AV},
and thus validates Eq.\ \ref{eq:SCAL3}.
Furthermore, since $P(t)\sim t^{-\theta}$ we find
$\xi(t)\simeq \frac{\theta}{\lambda}$ln $t$ at large $t$.
It follows that a persistent cluster of initial size $L$
has an average life-time
$\tau_{L}\sim exp(\frac{\lambda}{\theta}L)$ for large $L$.
The exponential dependence of the life-time of the cluster on its size
reflects the slow algebraic decay of persistence.

Our argument can be extended to show why $\tau$ and $z$ 
are possibly non-universal.
First of all, the exponent relation $\omega=\theta$ makes
it possible to write the formal relation 
$n(1,t)=g(\tau,n_{0})P(t)$\cite{EXP}.
Combined with Eq.\ \ref{eq:AV} and using the result 
$\overline l(t)=1/n_{0}$,
we obtain the relation $g(\tau,n_{0})+n_{0}=1$, which 
expresses implicitly the dependence of $\tau$ on $n_{0}$.
For instance, if Eq.\ \ref{eq:SCALFORM} were exact 
for {\it all} values of $k$, then $P(t)\approx
\int_{1}^{\infty}n(s,t)ds=\frac{n(1,t)}{\tau-1}$
so that $g(\tau,n_{0})\approx \tau-1$,
from which it follows that $\tau\simeq 2-n_{0}$. 
This result, although not exact, is consistent with 
the bounds $1<\tau<2$, and appears to be valid in
the high density limit $n_{0}\to 1$, as indicated
by the numerical values in Table I.

\myfigure{\epsfysize3.0in\epsfbox{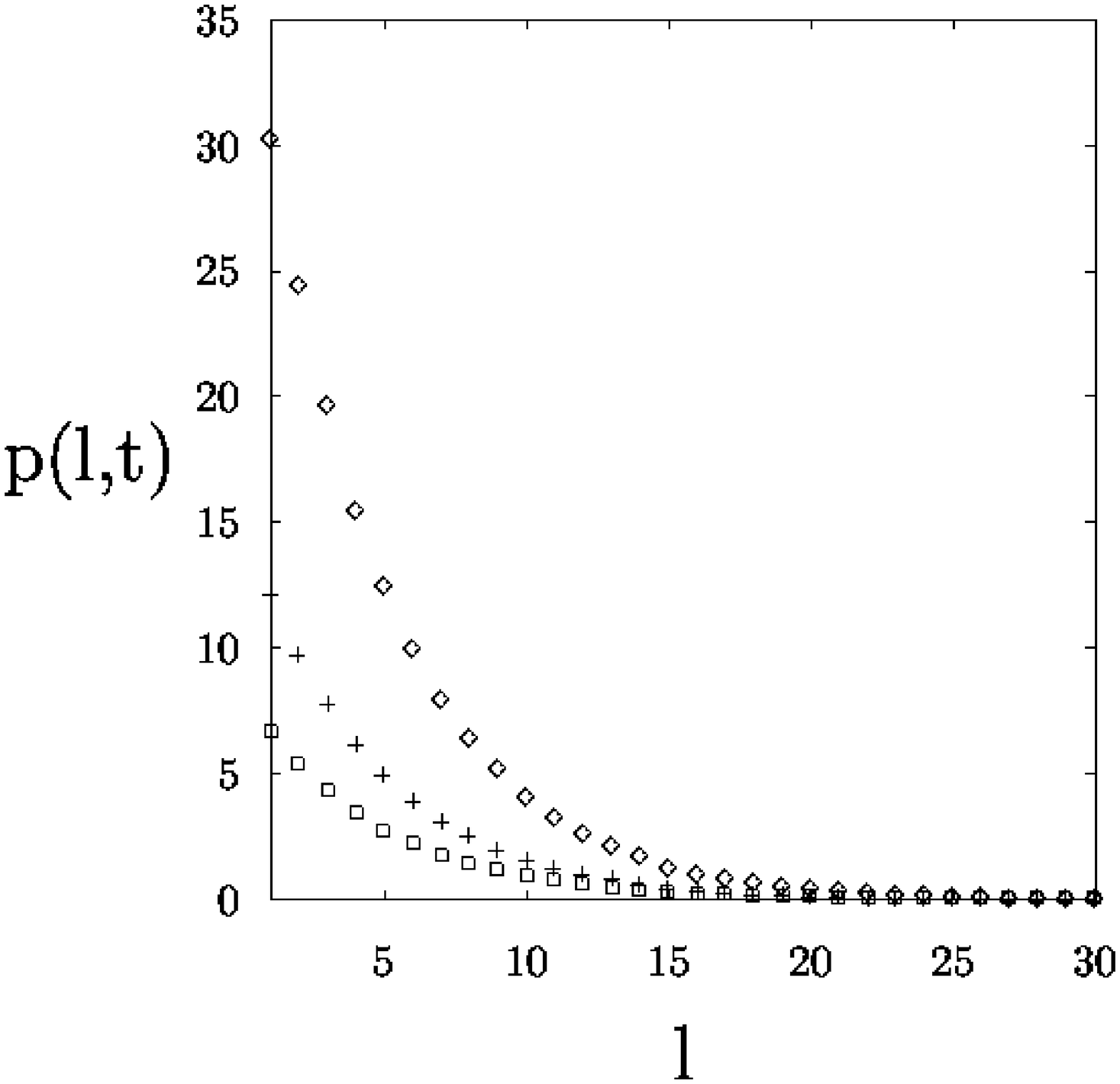}}
{\vskip-1.0in Fig.\ 4~~ 
The cluster size distribution $p(l,t)$ decays exponentially
with $l$ at all times. The figure shows $p(l,t)$ plotted 
against $l$ for $t=20$, 40 and 120 (top to bottom) and $n_{0}=0.2$. The vertical
separations have been enhanced for clarity.}

In summary, we have shown that the spatial distribution of persistent
clusters in one dimension exhibits rich dynamic scaling characterised by 
two new exponents. We have given rigorous arguments on the bounds
and universality properties of these exponents, which is
well-supported
by numerics. Interestingly,
the normalized size distribution of persistent clusters was found to be
{\it independent} of time.

Our work is the first study that brings out the non-trivial
features in the spatial distribution of persistent sites
in a one dimensional model.
The dynamic scaling form in Eq.\ \ref{eq:SCALFORM} is by no means
specific to the model studied, and we have observed similar forms 
in other one dimensional systems ---diffusion equation and kinetic
Ising model, for example.
Similar scaling in size distribution has 
been observed in entirely different contexts also---for instance,
diffusion-limited cluster aggregation\cite{FAMILY,VAN,TAKA}
and diffusion-limited deposition\cite{RACZ}.
The feature that is common to all these processes is the
irreversible coalescence of clusters (empty intervals in our case).

We are grateful to Satya Majumdar for discussions and for 
pointing out the similarities to aggregation models. G. M thanks
B. Derrida and P. R thanks D. Stauffer for critical reading
of the manuscript and valuable suggestions.

\begin{table}
\narrowtext
\begin{tabular}{cccc}
$n_{0}$ & 0.2 & 0.5 & 0.8\\
\hline
$\theta$ & 0.3718 $\pm$ 0.0001 & 0.3769 $\pm$ 0.0000 & 0.3729 $\pm$ 0.0001\\
$\omega$ & 0.3763 $\pm$ 0.0001 & 0.3767 $\pm$ 0.0000 & 0.3710 $\pm$ 0.0002\\
$z$ & 0.5766 $\pm$ 0.0005 & 0.5107 $\pm$ 0.0003 & 0.4392 $\pm$ 0.0005\\
$\tau$ & 1.3502 $\pm$ 0.0262 & 1.2596 $\pm$ 0.0357 & 1.1277 $\pm$ 0.0383\\
\end{tabular}
\caption {Exponents $\theta$, $\omega$, $z$ and $\tau$ as
measured from the simulations, for
three values of initial density $n_{0}$. The exact value of
$\theta$ is 0.375, independent of $n_{0}$. The $\omega$ values
presented correspond to $k=1$ and the $\tau$ values to $t=12000$.
The errors given are only statistical.}
\label{tab:TAB1}
\end{table}

\end{multicols}


\begin{references}
\bibitem{EXP} This follows from putting $n(1,t)=
g_{1}(n_{0})t^{-\theta}$, $P(t)=\sum_{k}n(k,t)=
g_{2}(\tau,n_{0})t^{-\theta}$ and then defining 
$g=g_{1}/g_{2}$.
\end{references}
\end{document}